\documentclass[a4paper, oneside, twocolumn, notitlepage, 10pt]{extarticle_ecoc}
\usepackage{ecoc}
\usepackage[justification=justified, format=plain]{caption} 
\makeatletter
\renewcommand\NAT@open{[}  
\renewcommand\NAT@close{]} 
\renewcommand\@cite[2]{\textsuperscript{\NAT@open#1\NAT@close}} 
\makeatother

\begin{document}
\selectlanguage{english}    


\title{Hybrid-Integrated Dual-Wavelength Laser Frequency Locked to an Integrated Coil-Resonator for Optical Fiber Sensing}%


\author{
    Mohamad Hossein Idjadi\textsuperscript{(1)}, Stefano Grillanda\textsuperscript{(1)},
    Nicolas Fontaine\textsuperscript{(1)}, Kaikai Liu\textsuperscript{(2)},
    Kwangwoong Kim\textsuperscript{(1)}, \\Tzu-Yung Huang\textsuperscript{(1)}, Cristian Bolle\textsuperscript{(1)}
    Rose Kopf\textsuperscript{(1)}, Mark Cappuzzo\textsuperscript{(1)}, and Daniel J. Blumenthal\textsuperscript{(2)}
}

\maketitle                  


\begin{strip}
    \begin{author_descr}

        \textsuperscript{(1)} Nokia Bell Labs, Murray Hill, NJ 07974, United States, 
        E-mail: mohamad.idjadi@nokia-bell-labs.com 

        \textsuperscript{(2)} University of California Santa Barbara, Department of Electrical and Computer Engineering, Santa Barbara, CA 93106, United States

    \end{author_descr}
\end{strip}

\renewcommand\footnotemark{}
\renewcommand\footnoterule{}


\begin{strip}
    \begin{ecoc_abstract}
        We demonstrate dual-frequency stabilization of a hybrid-integrated multi-channel laser to an integrated high Q-factor silicon nitride (SiN) coil resonator with more than 40 dB frequency noise suppression. The frequency locked channels are utilized for a proof-of-concept fiber sensing experiment.   \textcopyright2024 The Author(s)
    \end{ecoc_abstract}
\end{strip}


\section{Introduction}
\begin{figure*}[b!]
    \centering
    \includegraphics[height=5.8cm]{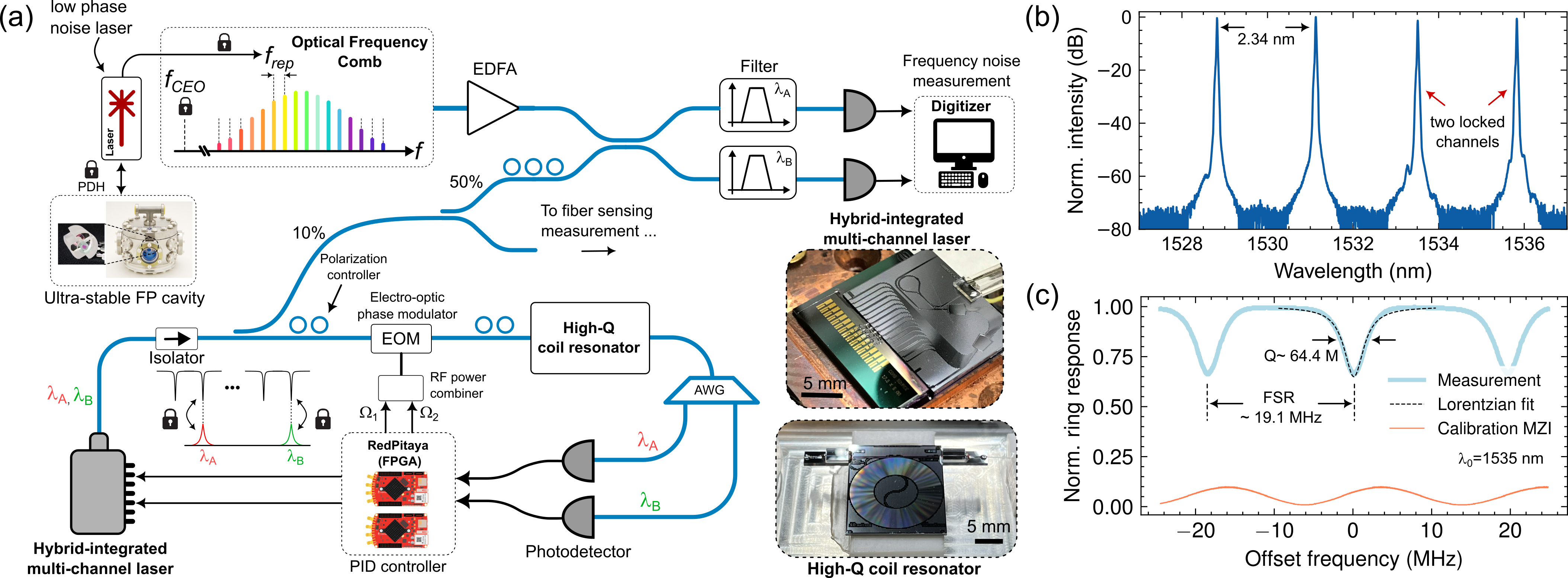}
    \caption{\textbf{(a)} Experimental setup used to PDH lock two channels of the hybrid integrated multi-channel laser to an integrated, SiN coil resonator (top part of the schematic shows also the setup for the frequency noise measurement). \textbf{(b)} Laser spectrum with four simultaneous active channels. \textbf{(c)} Normalized resonance response of the integrated SiN coil resonator. The micro-photo graph of the hybrid-integrated multi-channel laser and the coil resonator are shown in (a). PDH Pound-Drever-Hall, FSR free spectral range.}
    \label{fig1}
\end{figure*}

Low noise lasers with excellent long-term stability and high spectral purity are fundamental components of many applications, such as fiber sensing \cite{example:ofc_mazur, mecozzi2023use}, microwave photonics \cite{example:JLT_Morton, sun2024integrated, kudelin2024photonic} and optical atomic clocks \cite{example:RMP_atomic}. 
Moreover, having multiple laser wavelengths locked to a common reference cavity leads to enhanced sensing performance and opens up possibilities for further applications, such as dual-wavelength phase-sensitive optical time-domain reflectometry (OTDR) \cite{example:Opex_OTDR} and low-phase noise microwave generation \cite{liu2024low, example:Opex_microwave}. 
Ultra-narrow linewidth lasers have been developed extensively \cite{alkhazraji2023linewidth}, along with rapid advancements in stable and high quality-factor (Q-factor) optical cavities and resonators \cite{liu2024low, example:Optica_SiN_platform}. 
Despite recent advancements, developing integrated optical frequency technology in a CMOS-compatible process that offers multi-wavelength capability, milliWatt power per comb line, and ultra-narrow spectral linewidth remains challenging. 
We have developed a scalable and narrow-linewidth multi-channel laser technology with uniform milliwatt-level optical power per line through hybrid integration of a silica-on-silicon chips with III-V gain chips \cite{example:jlt_comb_source}. 
Moreover, our collaborators at UCSB have developed a CMOS compatible ultra low loss silicon nitride (SiN) platform with propagation loss of less than 0.5 dB/m across C-band that enables resonator with Q-factor >100 million in C-band and L-band \cite{liu2023integrated, example:Optica_SiN_platform}.

Here, we demonstrate simultaneous frequency locking of two wavelengths of the multi-channel hybrid-integrated laser source to the resonance frequencies of a high Q-factor SiN integrated coil resonator. 
Frequency locking of two free-running channels to the same reference frequency enables applications such as fiber sensing, where having stable and correlated frequencies is essential. 
We have demonstrated more than 4 orders-of-magnitude frequency noise suppression of both channels at low Fourier frequencies and about 3 orders-of-magnitude improvement in fractional frequency stability of the frequency difference between the two laser channels at gate time of 0.1 sec. 
As a proof-of-concept demonstration, the two frequency locked lines are used to sense small acoustic induced phase perturbations experienced by a 10-km fiber spool. 

\section{Integrated devices}
\begin{figure*}[t!]
    \centering
    \includegraphics[height=9.6cm]{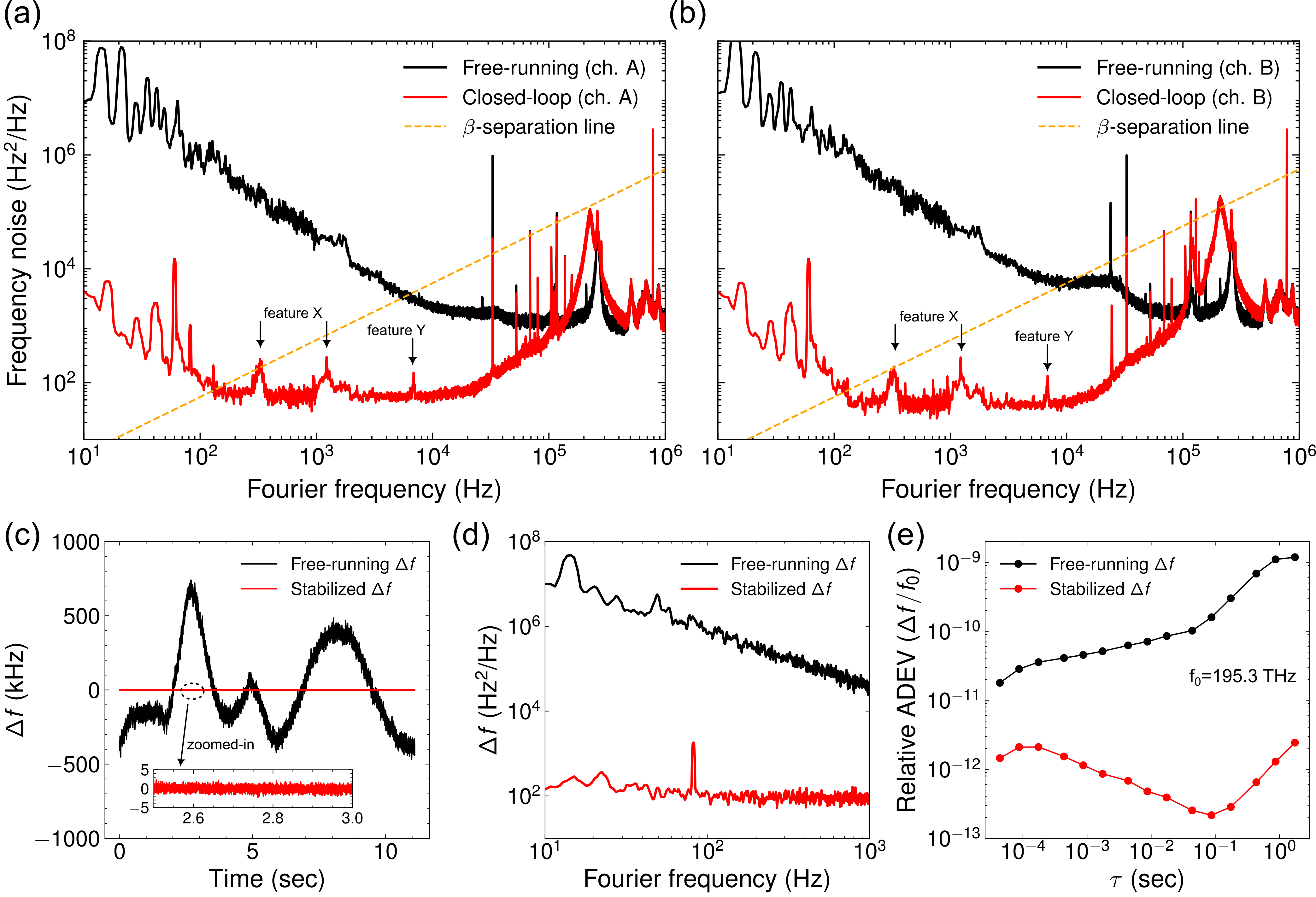}
    \caption{Dual-frequency stabilization and frequency noise measurement. \textbf{(a, b)} The frequency noise PSD of free-running and locked channels A and B, respectively. The highlighted features X, Y show up in the cross-spectrogram of the fiber sensing experiment (reported in Fig. \ref{fig3}). (a) and (b) share the same y-axis. \textbf{(c)} The difference in beat note frequencies between two laser channels ($\Delta f$) in time-domain. \textbf{(d)} The PSD of $\Delta f$. \textbf{(e)} The relative Allan deviation of $\Delta f$ is normalized to optical frequency of 195.3 THz. }
    \label{fig2}
\end{figure*}
Top-view photographs of the hybrid integrated multi-channel laser and of the integrated coil resonator are shown in the insets of Fig. \ref{fig1}(a). 
The hybrid integrated multi-channel laser is made by 4 III-V based reflective semiconductor optical amplifiers (RSOAs) and a passive silica-on-silicon chip. 
The RSOAs have highly reflective coatings on the back-facets and anti-reflection coatings on the front facets. 
The silica-on-silicon chip integrates an arrayed waveguide grating (AWG) and a Sagnac loop. 
The AWG pass-bands select a narrow band from the broad gain spectrum of the RSOAs and combine the light emitted by each RSOA on the same waveguide. 
Optical coupling between the RSOAs and the silica chip is achieved using ball lenses (two per channel). 
The laser cavities are formed between the RSOAs back-facets and the Sagnac loop. 
Figure \ref{fig1}(b) shows simultaneous lasing of four wavelengths of the laser where the two channels at wavelength of 1533.51 nm and 1535.83 nm are simultaneously frequency locked to the coil resonator; in this condition the optical power of each channel is about 7 mW. 
Additional details can be found in Ref. \cite{example:jlt_comb_source}. The assembled laser chip has a footprint of about 17.5 x 20 mm$^2$.    

The coil resonator has 10 m length, and was fabricated in a CMOS compatible, ultra-low loss SiN waveguide platform with 80 nm thickness and 6 $\mu$m width \cite{liu2023integrated, example:Optica_SiN_platform}. 
Figure \ref{fig1}(c) shows the normalized coil resonator transmission measured with a calibrated MZI with measured Q-factor of 64.4 million and a free spectral range of about 19.1 MHz at  wavelength of 1535~nm. 
The size of the 10 m coil resonator is 26 x 21~mm$^2$. 
Both the coil resonator and the laser are packaged in a temperature-controlled enclosure.

\section{Multi-channel laser frequency stabilization and frequency noise measurement}

Figure \ref{fig1}(a) shows the setup for multi-channel laser frequency stabilization using Pound-Drever-Hall locking technique \cite{drever1983laser}. The output of the hybrid-integrated laser is phase modulated using an electro-optic modulator (EOM). The EOM phase modulates both channels at 10 and 17 MHz, simultaneously, that are synthesized by low noise STEMlab Red Pitaya FPGAs. The output of the EOM is filtered by the high Q-factor coil resonator. An AWG is utilized to de-multiplex two channels ($\lambda_A$ and $\lambda_B$). The phase modulated optical signals are photodetected, amplified using transimpedance amplifiers, and digitized by the FPGAs. The error signal generated by the FPGAs modulate laser channels bias current to lock them to resonances of the SiN coil.

A small portion of the laser output is tapped out for frequency noise measurement and fiber sensing. A fully-stabilized octave-spanning fiber frequency comb, stabilized to an ultra-stable Fabry-P\'erot cavity, is mixed with the hybrid-integrated laser frequencies and the RF beat notes ($f_{b1}, f_{b2}$) are digitized and processed for frequency noise analysis\cite{idjadi2024modulation}. Figure \ref{fig2}(a), (b) show frequency noise power spectral density (PSD) of channels A and B, respectively, before and after frequency locking to two resonances of the optical frequency reference (\textit{i.e.} SiN resonator). As shown in Fig. \ref{fig2}(a) and (b), the frequency noise of free-running laser channels is suppressed by about 40 dB at low Fourier frequencies. The $\beta$-separation line suggests that the integral linewidth of free-running channel $\lambda_A$ and $\lambda_B$ is reduced from 74.4 kHz to 812 Hz and from 86.5 kHz to 791 Hz, respectively.

Figure \ref{fig2}(c) shows the beat note frequency difference between two laser channels (\textit{i.e.}, $\Delta f= f_{b1}-f_{b2}$) recorded over 10 sec and Fig. \ref{fig2}(d) reports the PSD of $\Delta f$. As plotted in Figs. \ref{fig2} (c) and (d), once two laser channels are frequency locked to the SiN coil resonator, their frequency difference fluctuation is significantly reduced. In Fig. \ref{fig2}(e), the Allan deviation is calculated for the beat note frequency difference normalized to optical frequency of 195.3 THz (wavelength of 1535 nm) where the fractional frequency stability of $\Delta f$ reaches to $2\times10^{-13}$ at 0.1 sec gate time. 
\section{Fiber sensing experiment}
\begin{figure*}[ht]
    \centering
    \includegraphics[height=7.8cm]{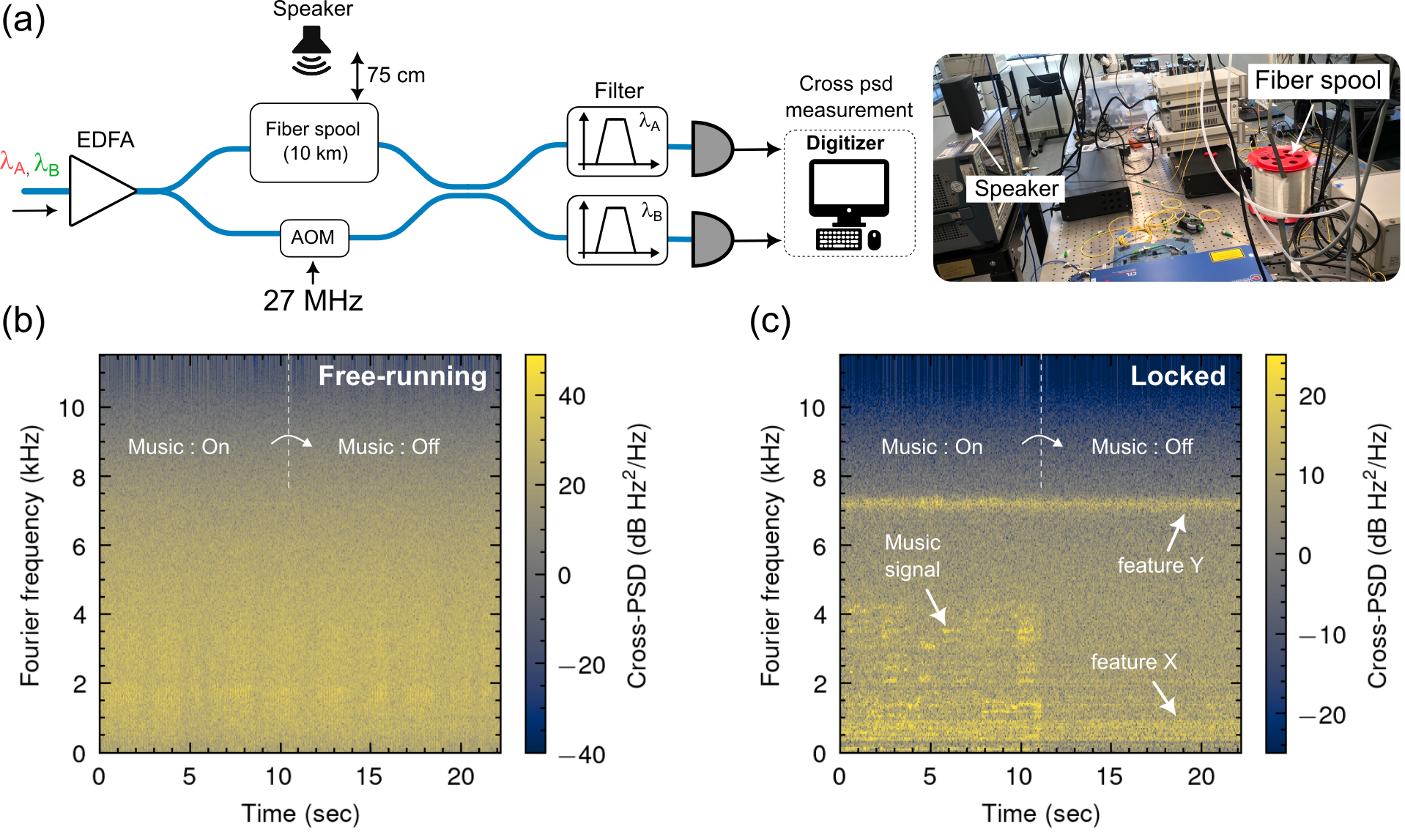}
    \caption{Fiber sensing experiment. \textbf{(a)} The measurement setup utilizing a 10 km fiber spool as an acoustic sensor. The cross-spectrogram of two \textbf{(b)}  free-running, and \textbf{(c)} frequency-locked laser channels are recorded over a span of 22 seconds. In both scenarios, music is played for the initial 11 seconds. }
    \label{fig3}
\end{figure*}

Figure \ref{fig3}(a) shows the fiber sensing setup where the optical signal with two wavelengths $\lambda_A$ and $\lambda_B$ is amplified and passed through a Mach-Zehnder interferometer with an acousto-optic modulator (AOM) in the bottom arm, that shifts frequency of both channels by 27 MHz, and a 10-km single mode fiber spool in another arm acting as the sensor. The two electric fields are mixed using a 2$\times$2 fiber coupler followed by optical band pass filters tuned to their corresponding wavelength channel. Optical signals are photodetected and digitized for further processing. 

As a proof-of-concept demonstration, we played a music for 11 seconds from a speaker at 75 cm away from the fiber spool with measured average sound intensity of 62 dB at the fiber sensor location. Figure \ref{fig3}(b) and (c) show the measured cross-spectrogram of two beat notes recorded for 22 sec while two laser channels are free-running and locked, respectively. As illustrated in Figs. \ref{fig3} (b,c), the music is played for the first 11 sec and then it is turned off for the rest of the experiment. 

As shown in Fig. \ref{fig3}(b), when two laser channels are free-running, the induced frequency/phase perturbation due to acoustic noise is masked by the phase noise of free-running laser channels. However, as shown in Fig. \ref{fig3}(c), there is a clear indication of acoustic phase perturbation for the first 11 sec (\textit{i.e.}, when the music is playing) when two laser channels are locked to a common optical frequency reference (\textit{i.e.}, coil resonator). Moreover, there are frequency features (X, Y) in the locked cross-spectrogram at approximately 0.5-1 kHz and 7 kHz Fourier frequencies that match very well with measured frequency noise PSD of each frequency stabilized channel in Fig. \ref{fig2}(a,b).   


\section{Conclusions}
Two channels of a hybrid-integrated multi-channel laser are frequency-locked to resonances of a high Q-factor integrated SiN coil resonator. Through simultaneous frequency locking of the channels of the integrated laser to the same optical cavity, as a proof-of-concept, we have demonstrated a significant improvement in sensitivity for detecting acoustic signals. Combining the scalable multi-channel laser technology with the ultra-low-loss SiN platform represents a promising step forward in enabling precision optical interferometry, with a substantial enhancement in sensing capabilities.

\clearpage
\section{Acknowledgements}
We wish to thank Mikael Mazur for fruitful discussions.




\vspace{-4mm}

\end{document}